\documentclass{preprint}
\usepackage{ncg,qft}
\renewcommand*{\bar}[1]{\overline{#1}}
\renewcommand*{\tilde}[1]{\widetilde{#1}}
\newcommand*{\sqrthalf}{\frac{1}{\sqrt{2}}}
\newcommand*{\half}{\frac{\scriptstyle 1}{\scriptstyle 2}}
\newcommand*{\MPlanck}{m_\mathrm{p}{}}
\newcommand*{\LPlanck}{l_\mathrm{p}{}}
\newcommand*{\phibar}{\bar{\phi}}
\newcommand*{\psibar}{\bar{\psi}}
\newcommand*{\diag}[1]{\mathop{\mathrm{diag}}\nolimits(#1)}
\newcommand*{\vev}[1]{\langle #1 \rangle}

\title{Path Integral Quantisation of Finite Noncommutative Geometries}
\author{Mark Hale\thanks{e-mail: mark.hale@physics.org}\\
Department of Mathematical Sciences,\\
Science Laboratories,\\
University of Durham,\\
Durham, DH1 3LE, UK}
\date{May 28, 2001}

\begin{document}
\maketitle

\begin{abstract}
We present a path integral formalism for quantising gravity
in the form of the spectral action.
Our basic principle is to sum over all Dirac operators.
The approach is demonstrated on two simple finite noncommutative
geometries: the two-point space, and the matrix geometry $\Matrix{2}{\Cset}$.
On the first, the graviton is described by a Higgs field, and on the
second, it is described by a gauge field.
We start with the partition function and calculate the propagator and
Greens functions for the gravitons.
The expectation values of distances are evaluated, and we
discover that distances shrink with increasing graviton excitations.
We find that adding fermions reduces the effects of the
gravitational field.
A comparison is also made with Rovelli's canonical quantisation
approach, and with his idea of spectral path integrals.
We include a brief discussion on the quantisation of a Riemannian manifold.
\end{abstract}

\section{Introduction}

One of the greatest successes of noncommutative geometry has been
the unification of the forces of nature into a single gravitational
action---the spectral action \cite{Connes:spectral,nonperturb_spectral}.
This has been achieved at the classical level for an Euclidean signature.
It does this by using the Kaluza-Klein idea of rewriting all the gauge
fields as components of a metric on a more structured spacetime.
Noncommutative geometry succeeds where Kaluza-Klein fails as it is
not limited to ordinary differential manifolds.
For introductions to noncommutative geometry see
\cite{Schucker:intro,Varilly:intro,Landi:intro}.

The question of how to quantise a field theory on a general
noncommutative geometry remains largely unresolved.
Conventional techniques work on Riemannian-like manifolds and have
been used on noncommutative extensions, such as almost commutative
geometries (the tensor product of a Riemannian manifold with a
finite noncommutative geometry) and the noncommutative torus
\cite{Krajewski:nct}.
Beyond this, most efforts have focused on quantising a particular
noncommutative geometry \cite{Mangano,Heller_Sasin:groupoid,Heller_Sasin:comp}.

In this paper, we present a path integral approach that is
applicable to any noncommutative geometry.
It has been developed to quantise the spectral action,
which is the natural geometric action for a noncommutative geometry.
The Dirac operator is the dynamical variable of the spectral action,
and plays the role of the metric.
A path integral should therefore be some sort of
``sum over Dirac operators".
We define what this might mean by appealing to the conventional
path integral formalism.
Our approach builds on, and complements, the work done by
Rovelli in \cite{Rovelli:ncg}.

The outline of this paper is as follows.
We begin in section \ref{sec:formalism} with a detailed description
of our path integral formalism.
Then, in section \ref{sec:twopt}, we apply it to the two-point space,
and, in section \ref{sec:matrix}, to the matrix geometry $\Matrix{2}{\Cset}$.
For these geometries, the path integrals are standard
(finite dimensional) integrals,
so the technical difficulties associated with functional integration are avoided.
To keep the examples clear and concise, we restrict ourselves to
$(\algebra,\hilbert,D)$ spectral triples.
That is, we ignore real structure, orientability and Poincar\'{e} duality,
which do not play an essential role in the discussion.
As such, our example geometries can be considered as fragments of
larger noncommutative geometries that do conform to all the axioms
set out in \cite{Connes:axioms}.

In section \ref{sec:rovelli}, we make a comparison with the
canonical quantisation approach taken in \cite{Rovelli:ncg}.
The results we obtain from using our approach to quantise
the geometry used in \cite{Rovelli:ncg} are given in
section \ref{sec:rovelli-geom}.
An idea for a path integral approach is also proposed in \cite{Rovelli:ncg}.
In section \ref{sec:spectral}, we highlight the differences between
this approach and our path integrals.
We follow this, in section \ref{sec:riemann}, with a brief
discussion on the quantisation of a Riemannian manifold.
Section \ref{sec:conclusion} marks the end, with the conclusion.

Note: we work with an Euclidean signature, i.e.\ Riemannian means
Riemannian not pseudo-Riemannian.

\section{Path Integral Quantisation}
\label{sec:formalism}

We decided to work on a path integral approach, rather than a
canonical approach, because it requires knowledge of only the fields,
and not their dynamics.
To be able to canonical quantise a noncommutative geometry,
we would need a general procedure for finding the phase space,
and constructing a symplectic structure on it.
Conventionally, this amounts to finding the canonical momenta and
using the Poisson bracket.
In contrast, path integrals need a (gauge invariant) measure
on the space of histories.
Deciding how to parameterise this space is thus an important
consideration.
The advantage lies in that this does not depend on the details of
the action, unlike finding the phase space.
The only things that really matter are the fields, because they
determine the measure.
One of the other benefits of using path integrals is they are
explicitly covariant.

A good starting point for developing a path integral formalism for
noncommutative geometry is the conventional formalism.
It has lead to standard model predictions that agree spectacularly
with experiment, so it should be incorporated as a special case.
Since the standard model action can be expressed in the form of a
spectral action, a dictionary can be set up between
noncommutative geometry and quantum field theory.
This makes it apparent that the (gauge) fields parameterise the Dirac operator.
So, the space of histories of the fields is equivalent to the
space of histories of the Dirac operator.
From the noncommutative geometry point of view then, the degrees of
freedom of the Dirac operator correspond to the fields in the spectral
action, and hence give the path integration measure.
Thus, in principle, we can path integral quantise a general spectral
action.
Schematically, the general partition function can be written as
\begin{equation}
Z := \int\!\mathcal{D}D\, \e^{-\Tr f\left(D^2/\Lambda^2\right)},
\end{equation}
where $D$ is the Dirac operator.
The function $f$ and parameter $\Lambda$ are the cutoffs for the
spectral action.

\section{The Two-Point Space and Higgs Gravity}
\label{sec:twopt}

The two-point space is the simplest example of a
noncommutative geometry.
It consists of just two points which we label $L$ and $R$.
The spectral triple is given by
\begin{eqnarray}
\algebra & := & \Cset\oplus\Cset = \left\{
f := \left(\begin{array}{cc}
f_L & 0 \\
0 & f_R \end{array}\right)\right\}, \nonumber \\
\hilbert & := & \Cset\oplus\Cset, \\
D & := & \frac{1}{\hbar}\left(\begin{array}{cc}
0 & m \\
\bar{m} & 0 \end{array}\right), \nonumber
\end{eqnarray}
where $m$ is a complex constant which fixes the distance between
the two points.
It can almost be made into a real spectral triple;
there is an obvious grading $\Gamma := \diag{1, -1}$
and a real structure $J$ given by complex conjugation.
However, they do not satisfy all of Connes' axioms.
The two-point space can best be described as a ``scaled" even Fredholm module.

Some may be unsettled by the appearance of $\hbar$ in the Dirac
operator before quantisation.
It is used only to follow the convention that $m$ has units of
mass, rather than inverse length, and so can be omitted.
Alternatively, one could view $\hbar$ as the noncommutative geometry
version of $c$.
In the same way that $c$ relates space and time on a Lorentzian
manifold, $\hbar$ relates space and (inverse) mass on a noncommutative
geometry (``\textit{spacemass\/}").
No $\hbar$ is required for quantisation as the spectral action is naturally
dimensionless.
We, however, will take our actions to have the usual dimensions of $\hbar$.

To move from a static (flat) space to a dynamic (curved) space, we
promote the constant $m$ to a variable $\phi$, which will play the
role of the gravitational field.
This is the analogue of moving from $\eta_{\mu\nu}$ to $\guv(x)$
on a Lorentzian manifold.
In fact, $\phi$ is really a connection, so it plays the role of a
vierbein/spin connection rather than a metric.
In the context of the standard model, $\phi$ is interpreted as the
Higgs field, hence we refer to this as Higgs gravity.

The spectral action is taken to be
\begin{equation}
S := \frac{1}{G}\Tr D^2 = \frac{2\LPlanck^2}{\hbar}|\phi|^2,
\end{equation}
where $G$ is the gravitational coupling constant,
and $\LPlanck := 1/\sqrt{\hbar G}$ is the Planck length.
It has a U(1) symmetry which comes from $\Inn(\algebra)$, the
inner automorphism group of $\algebra$.
For the two-point space, $\Inn(\algebra) \cong \Ugrp(1)\times\Ugrp(1)$,
which acts on $\phi$ via the U(1) transformations given by the homomorphism
$\Ugrp(1)\times\Ugrp(1) \to \Ugrp(1) : (g,h) \to gh^{-1}$.
The inner automorphisms are analogous to the diffeomorphisms of
general relativity.
They are often referred to as internal diffeomorphisms.

Varying the action, the equations of motion are simply
\begin{equation}
\phi = 0 ,\quad \phibar = 0.
\end{equation}
Using Connes' distance formula,
\begin{equation}
d(x,y) := \sup_{f\in\algebra} \big\{\big|\bra{x}f\ket{x}-\bra{y}f\ket{y}\big| : ||[D,f]|| \leq 1\big\},
\label{eq:distance_formula}
\end{equation}
the distance between the two points is
\begin{equation}
d(L,R) = \sup_{f\in\algebra} \big\{\big|f_L-f_R\big| : \frac{|\phi|^2}{\hbar^2}|f_L-f_R|^2 \leq 1\big\}
= \frac{\hbar}{|\phi|} = \frac{\MPlanck}{|\phi|}\LPlanck,
\label{eq:twopt-distance}
\end{equation}
where $\MPlanck$ is the Planck mass.
So, classically, the metric structure $D$ vanishes and the
distance is infinite.

Now, we quantise by doing path integrals over
$\phi$ and $\phibar$, the degrees of freedom of $D$.
The partition function is thus
\begin{equation}
Z := \int\!\d D\, \e^{-S/\hbar}
= \int\!\d\phibar\,\d\phi\, \exp\left(-\frac{2|\phi|^2}{\MPlanck^2}\right).
\end{equation}
Since the action has a U(1) symmetry, we shall employ some gauge-fixing.
This involves nothing more than switching to polar coordinates $(r,\theta)$,
and dropping the irrelevant $\theta$ integration.
Note: as the number of gauge degrees of freedom is finite,
gauge-fixing is not strictly necessary
(the $\theta$ integration does not give an infinite contribution).
After integrating out gauge equivalent Dirac operators, the partition function
reduces to
\begin{equation}
Z = \int\limits_0^\infty\!\d\phi\, \phi\exp\left(-\frac{2\phi^2}{\MPlanck^2}\right)
= \frac{\MPlanck^2}{4},
\end{equation}
where $\phi$ is now used to denote the positive real field $|\phi|$.

Expectation values are calculated in the usual fashion.
For example,
\begin{eqnarray}
\vev{\phi} & = & \frac{1}{Z}\int\limits_0^\infty\!\d\phi\, \phi^2 \exp\left(-\frac{2\phi^2}{\MPlanck^2}\right)
= \frac{\sqrt{2\pi}}{4}\MPlanck, \\
\vev{d(L,R)} & = & \frac{1}{Z}\int\limits_0^\infty\!\d\phi\, \MPlanck\LPlanck \exp\left(-\frac{2\phi^2}{\MPlanck^2}\right)
= \sqrt{2\pi}\,\LPlanck.
\label{eq:twopt-d0}
\end{eqnarray}
Here, we see that in the vacuum state, $\phi$ has acquired a v.e.v.,
and the distance has become finite.
Though, the classical distance relation \eqnref{eq:twopt-distance}
no longer holds.

In general,
\begin{equation}
\int\limits_0^\infty\!\d\phi\, \phi^n \exp\left(-\frac{2\phi^2}{\MPlanck^2}\right)
= \frac{1}{2}\Gamma\left(\frac{\scriptstyle n+1}{\scriptstyle 2}\right)
\left(\frac{\MPlanck}{\sqrt{2}}\right)^{n+1}.
\end{equation}
Thus, the Greens functions are
\begin{equation}
\vev{\phi^n}
= \Gamma\left(\frac{\scriptstyle n+2}{\scriptstyle 2}\right)
\left(\frac{\MPlanck}{\sqrt{2}}\right)^n.
\end{equation}
In particular, the propagator functions can be expressed as
\begin{equation}
\vev{(\phi\phi)^n} = n!\left(\frac{\MPlanck^2}{2}\right)^n
\end{equation}
for $n\in\Zset$.
These reproduce the usual propagator combinatorics
(i.e.\ Wick contractions) for a complex scalar field.

In an excited state, the distance $d(L,R)$ is given by
its expectation value in a background of propagators.
So, for the $N$th particle state,
\begin{equation}
\vev{d(L,R)}_N = \frac{1}{Z_N}\vev{\phi^N d(L,R) \phi^N},
\end{equation}
where $Z_N = \vev{(\phi\phi)^N}$.
This evaluates to
\begin{equation}
\vev{d(L,R)}_N = \frac{\Gamma(N+\half)}{\Gamma(N+1)} \sqrt{2}\,\LPlanck.
\end{equation}
The distance thus gets successively smaller as the number of gravitons
(Higgs particles) is increased.
Using Stirling's formula, we find that the distance shrinks to zero
in the $N \to \infty$ limit, and so the two points merge into one.
The metric $D$ correspondingly becomes infinite, since the description
of the geometry as two points is no longer valid.
This resembles the behaviour of a high curvature limit,
i.e.\ gravitational collapse to a black hole.

The spectral action can be supplemented with the fermionic term
\begin{equation}
S_F := \langle\psibar,D\psi\rangle
= \psibar_L\phi\psi_R+\psibar_R\phibar\psi_L,
\label{eq:twopt-fermion}
\end{equation}
which is invariant under the full $\Ugrp(1)\times\Ugrp(1)$
symmetry.
Note that this is purely an interaction term---the fermions are
fixed at the points and do not propagate.
The equations of motion for gravity coupled to matter are
\begin{equation}
\begin{array}{lll}
\displaystyle
\frac{2\phi}{\MPlanck^2}+\psibar_R\psi_L=0, & \phibar\psi_L=0, & \phi\psi_R=0, \\
\\
\displaystyle
\frac{2\phibar}{\MPlanck^2}+\psibar_L\psi_R=0, & \psibar_R\phibar=0, & \psibar_L\phi=0.
\end{array}
\end{equation}
So, $\phi=0$, $\phibar=0$ and either $\psi_L=0$, $\psibar_L=0$
or $\psi_R=0$, $\psibar_R=0$.

Quantising as before, we write down the partition function,
\begin{equation}
Z = \int\!\d\phibar\,\d\phi\,\d\psibar\,\d\psi\,
\exp\left( -\frac{2|\phi|^2}{\MPlanck^2} -\langle\psibar,D\psi\rangle\right).
\end{equation}
Remember that the Hilbert space is complex, and not Grassmann, so
\begin{equation}
Z = \int\!\d\phibar\,\d\phi\, \frac{1}{\det D} \exp\left(-\frac{2|\phi|^2}{\MPlanck^2}\right)
= -\int\limits_0^\infty\!\d\phi\, \frac{1}{\phi} \exp\left(-\frac{2\phi^2}{\MPlanck^2}\right)
= \infty.
\end{equation}

This makes the v.e.v.\ $\vev{d(L,R)}$ ill-defined, while both
$\vev{\phi}$ and the propagator $\vev{\phi\phi}$ will be zero.
For the excited states ($N \geq 1$), the expectation values
continue to be well-behaved.
The effect of the fermions is to shield out the gravitational field,
by lowering the states by one.
If we were to take the tensor product of the Hilbert space with a
spinor Hilbert space $L^2(\spin(M))$, then the fermions would
enhance the gravitational field, by raising the states.

Note: for a generic finite noncommutative geometry, the fermion
contribution will be $(\det D)^{-k}$ where $k$ is the number of
fermion generations fixed by the Hilbert space.

\section{Matrix Geometries and Gauge Gravity}
\label{sec:matrix}

Next, we look at the quantisation of the simplest matrix geometry, $\Matrix{2}{\Cset}$.
Its spectral triple is
\begin{eqnarray}
\algebra & := & \Matrix{2}{\Cset} = \left\{
f := \left(\begin{array}{cc}
f_1 & f_2 \\
f_3 & f_4 \end{array}\right)\right\}, \nonumber \\
\hilbert & := & \Matrix{2}{\Cset}, \\
D & := & \frac{1}{\hbar}\left(\begin{array}{cc}
A_1 & A_2 \\
\bar{A}_2 & -A_1 \end{array}\right), \nonumber
\end{eqnarray}
where $D$ is an SU(2) gauge field, with $A_1$ real and $A_2$ complex.
This is a reduction of the even spectral triple obtained by
tensoring the representation with the Clifford algebra $\Cset\mathrm{l}(\Rset^2)$,
\begin{eqnarray}
\algebra' & := & \algebra, \nonumber \\
\hilbert' & := & \hilbert\oplus\hilbert \quad\mbox{with $f' := f \otimes \identity_2$}, \nonumber \\
D' & := & D \otimes \left(\begin{array}{cc}
0 & 1 \\
1 & 0 \end{array}\right), \\
\Gamma' & := & \identity_2 \otimes \diag{1, -1}. \nonumber
\end{eqnarray}
Moreover, this itself is the point-reduction of the real spectral triple
with $\algebra'' := C^\infty(\Rset^2)\otimes\algebra$,
$\hilbert'' := L^2(\spin(\Rset^2))\otimes\hilbert$
and $D'' := -\im\gamma^\mu(\partial_\mu+\im A_\mu)$.
The $C^*$-algebra $\Matrix{2}{\Cset}$ can be understood as being that of the
fuzzy sphere $\Sset^2_{(n=1)}$,
which only has the north and south poles as distinguishable points.

The spectral action evaluates to
\begin{equation}
S := \frac{1}{G}\Tr D^2
= \frac{2\LPlanck^2}{\hbar}\left(A_1{}^2+|A_2|^2\right),
\end{equation}
which is invariant under SU(2) gauge transformations.
Like the two-point space, the inner automorphisms
$\Inn(\algebra) \cong \Ugrp(2)$ act on $D$ via a homomorphism,
$\Ugrp(2) \to \SUgrp(2)$.
The homomorphism removes the trivial U(1) factor that
commutes with $D$.

As before, we shall quantise by first gauge-fixing the action.
This is most easily accomplished by changing to
spherical polar coordinates.
So, after dropping irrevelant factors, the partition function reads
\begin{equation}
Z = \int\limits_0^\infty\!\d\phi\, \phi^2\exp\left(-\frac{2\phi^2}{\MPlanck^2}\right)
= \frac{\sqrt{2\pi}}{16}\MPlanck^3,
\end{equation}
where $\phi := \sqrt{A_1{}^2+|A_2|^2}$.
Effectively, we have chosen a gauge-fixing condition such that
\begin{equation}
D = \frac{1}{\hbar}\left(\begin{array}{cc}
0 & \phi \\
\phi & 0 \end{array}\right).
\end{equation}
This gauge can be obtained from any other by performing
an SU(2) gauge transformation
\begin{equation}
D \to uDu^\dagger = D+u[D,u^\dagger]
\end{equation}
with
\begin{equation}
u = \frac{1}{2\sqrt{\phi(\phi-A_1)}}\left(\begin{array}{cc}
\phi-A_1+\bar{A}_2 & \phi-A_1-A_2 \\
-(\phi-A_1-\bar{A}_2) & \phi-A_1+A_2 \end{array}\right).
\end{equation}
The Greens functions for $\phi$ are
\begin{equation}
\vev{\phi^n}
= \frac{2}{\sqrt{\pi}}\Gamma\left(\frac{\scriptstyle n+3}{\scriptstyle 2}\right)
\left(\frac{\MPlanck}{\sqrt{2}}\right)^n.
\end{equation}
As one would expect, they reflect the combinatorics of a field
that can propagate through either a real mode ($A_1 \to A_1$)
or a complex one ($A_2 \to \bar{A}_2$).

The distance between the poles of the fuzzy sphere,
\begin{equation}
d(1,4) = \sup_{f\in\algebra} \big\{\big|f_1-f_4\big| : ||[D,f]|| \leq 1\big\},
\end{equation}
is not as straightforward to calculate as the distance between the
points of the two-point space.
Evaluating the condition $||[D,f]|| \leq 1$ gives
\begin{equation}
\frac{\hbar}{\phi} \geq \Bigg\{\:{
\big|(f_1-f_4)+(f_2-f_3)\big| \atop
\big|(f_1-f_4)-(f_2-f_3)\big|}
\quad\mbox{depending on which is larger.}
\end{equation}
This can be simplified by expressing it in terms of ``distances" and phases,
\begin{equation}
\frac{\hbar}{\phi} \geq \big|d_{14}\e^{\im\alpha}\pm\,d_{23}\e^{\im\beta}\big|,
\end{equation}
where $d_{14}\e^{\im\alpha} := (f_1-f_4)$ and $d_{23}\e^{\im\beta} := (f_2-f_3)$.
Squaring up both sides, it is then easy to determine the larger lower bound,
\begin{eqnarray}
\frac{\hbar^2}{\phi^2} & \geq & d_{14}{}^2\pm2d_{14}d_{23}\cos(\alpha-\beta)+d_{23}{}^2 \nonumber \\
& \geq & d_{14}{}^2+2d_{14}d_{23}|\cos(\alpha-\beta)|+d_{23}{}^2.
\end{eqnarray}
Hence, the upper bound on $d_{14}$ is
\begin{equation}
d_{14} \leq
-d_{23}|\cos(\alpha-\beta)|+\sqrt{\frac{\hbar^2}{\phi^2}-d_{23}{}^2\sin^2\theta}.
\end{equation}
Taking the supremum, the distance is therefore
\begin{equation}
d(1,4) = \frac{\hbar}{\phi} = \frac{\MPlanck}{\phi}\LPlanck.
\end{equation}
Similarly, we also find
\begin{equation}
d(2,3) = \frac{\hbar}{\phi} = \frac{\MPlanck}{\phi}\LPlanck.
\end{equation}
(We should clarify that there are no states $\bra{\psi_2}f\ket{\psi_2} = f_2$
and $\bra{\psi_3}f\ket{\psi_3} = f_3$,
but there are two (pure) states $\ket{\psi_2}$ and $\ket{\psi_3}$ such that
$|\bra{\psi_2}f\ket{\psi_2}-\bra{\psi_3}f\ket{\psi_3}| = |f_2-f_3|$.)

The expectation value of the distances, in the $N$th particle state, is
\begin{equation}
\vev{d}_N = \frac{\Gamma(N+1)}{\Gamma(N+\frac{\scriptstyle 3}{\scriptstyle 2})} \sqrt{2}\,\LPlanck.
\end{equation}
Just like the two-point space, the distances shrink to zero
in the $N \to \infty$ limit.
However, the nature of this collapse is rather different.
The K-groups of the fuzzy sphere do not change as it collapses to a point,
indeed $K_*(\Matrix{2}{\Cset}) \cong K_*(\Cset)$.
Whereas this is not the case for the two-point space, for which
$K_*(\Cset\oplus\Cset) \cong K_*(\Cset)\oplus K_*(\Cset)
\not\cong K_*(\Cset)$.
So, the collapse of the fuzzy sphere involves a change in
commutativity, rather than topology.

From a K-theory perspective, the fuzzy sphere is more like a
(noncommutative) point than a sphere.
It is referred to as a sphere because of its SU(2) symmetry.
In fact, the space of pure states of $\Matrix{2}{\Cset}$ is a 2-sphere.
Incidentally, the K-groups of a 2-sphere are actually isomorphic
to those of the two-point space.

The fermion action for the fuzzy sphere is
\begin{eqnarray}
S_F & := & \Tr\Psi^\dagger D\Psi \nonumber \\
& = & \psibar_1 A_1\psi_1+\psibar_2 A_1\psi_2-\psibar_3 A_1\psi_3-\psibar_4 A_1\psi_4 \nonumber \\
&& {}+\psibar_1 A_2\psi_3+\psibar_3\bar{A}_2\psi_1+\psibar_2 A_2\psi_4+\psibar_4\bar{A}_2\psi_2.
\end{eqnarray}
It contains twice as many fermions as \eqnref{eq:twopt-fermion}
due to the larger Hilbert space.
The contribution to the partition function will thus be
$(\det D)^{-2} = \phi^{-4}$.
This will have the effect of lowering the states by two.

\section{Comparison with Rovelli's Canonical Quantisation}
\label{sec:rovelli}

We tried to compare our path integral approach with Rovelli's
canonical approach (see \cite{Rovelli:ncg} for details),
but found problems with the example he gives.
Rovelli modified the spectral action in an effort to obtain
non-trivial equations of motion.
After careful examination, we found this actually had the opposite effect.
The action in question is
\begin{eqnarray}
S & := & \frac{1}{2}\Tr D\tilde{M}D \nonumber \\
& = & \frac{1}{2G}\left(\bar{m}_1 m_1+\e^{-\im\theta}\bar{m}_1 m_2+\e^{\im\theta}\bar{m}_2 m_1+\bar{m}_2 m_2\right).
\label{eq:rovelli-action}
\end{eqnarray}
But, this can be factorised as
\begin{eqnarray}
& = & \frac{1}{2G}\left(\bar{m}_1+\e^{\im\theta}\bar{m}_2\right)\left(m_1+\e^{-\im\theta}m_2\right) \hspace{1.78cm} \nonumber \\
& = & \frac{|m|^2}{2G},
\label{eq:rovelli-myaction}
\end{eqnarray}
where $m := m_1+\e^{-\im\theta}m_2$.
We thus end up with a much simpler action and set of equations of motion.
Canonical quantisation in this variable is a very different problem
from the one considered by Rovelli.

Physically, the interaction terms in \eqnref{eq:rovelli-action} allow
the particles $m_1$ and $m_2$ to spontaneously change into one another.
This is like a mixing term, so $m_1$ and $m_2$ will not make good
eigenstates.
As we have seen in \eqnref{eq:rovelli-myaction}, the linear combination
given by $m$ will make a good eigenstate.

Although the action \eqnref{eq:rovelli-action} is not spectral \textit{per se},
we can in fact still quantise it with our path integral approach.
We begin by rewriting the action in terms of an effective Dirac
operator, $D'$, so it is spectral:
\begin{equation}
S = \frac{1}{2}\Tr D\tilde{M}D
= \frac{1}{2}\Tr D^\dagger P^\dagger PD
= \frac{1}{2G}\Tr D'^\dagger D'
\end{equation}
Solving $P^\dagger P = \tilde{M}$ gives
\begin{equation}
P = \frac{1}{\sqrt{2G}}\left(\begin{array}{ccc}
1 & \e^{-\im\theta} & 0 \\
1 & \e^{-\im\theta} & 0 \\
0 & 0 & 0 \end{array}\right),
\end{equation}
thus
\begin{equation}
D' = \sqrt{G}PD = \sqrthalf\left(\begin{array}{ccc}
0 & 0 & m \\
0 & 0 & m \\
0 & 0 & 0 \end{array}\right).
\end{equation}
Further, a self-adjoint operator $D''$ can be constructed by
\begin{equation}
D'^\dagger D' = \left(\frac{D'^\dagger+D'}{\sqrt{2}}\right)^2
= D''^2 = \frac{1}{4}\left(\begin{array}{ccc}
0 & 0 & m \\
0 & 0 & m \\
\bar{m} & \bar{m} & 0 \end{array}\right)^2,
\end{equation}
since $D'$ is nilpotent.
The degrees of freedom of $D''$ are $m$ and $\bar{m}$,
just as we have proposed.
Quantising this, we end up with path integrals equivalent to those
for the two-point space.

The problem with trying to canonically quantise spectral actions for
finite noncommutative geometries is they have no phase space as such.
This could be taken to mean that they simply cannot be quantised,
but we have shown otherwise using path integrals.
Perhaps some generalisation of phase space is needed
(like tangent groupoids, see \cite[sec.~6]{Varilly:intro}),
or maybe the path integral approach is just more fundamental.

We could try to reverse-engineer our path integral approach
and find the canonical equivalent.
The Fourier mode expansion of the Higgs graviton, in terms of creation
and annihilation operators, should be
\begin{equation}
\hat{\phi} = \frac{\MPlanck}{\sqrt{2}} \left(\hat{a}+\hat{b}^\dagger\right) ,\quad
\hat{\phi}^\dagger = \frac{\MPlanck}{\sqrt{2}} \left(\hat{a}^\dagger+\hat{b}\right).
\end{equation}
Thus, the conjugate momentum operator should be
\begin{equation}
\hat{\pi} = \frac{\im}{\sqrt{2}\,\MPlanck} \left(\hat{a}^\dagger-\hat{b}\right) ,\quad
\hat{\pi}^\dagger = \frac{\im}{\sqrt{2}\,\MPlanck} \left(\hat{a}-\hat{b}^\dagger\right).
\end{equation}
What, then, does the classical quantity $\pi$ correspond to?
We leave the further exploration of these ideas for another time.

\section{Path Integral Quantisation of Rovelli's Geometry}
\label{sec:rovelli-geom}

Having quantised Rovelli's modified spectral action \eqnref{eq:rovelli-action}
using path integrals, we shall now do the same for the
un-modified spectral action
\begin{equation}
S := \frac{1}{G}\Tr(D+JDJ^{-1})^2,
\end{equation}
where
\begin{equation}
D := \frac{1}{\hbar}\left(\begin{array}{ccc}
0 & 0 & \phi_1 \\
0 & 0 & \phi_2 \\
\bar{\phi}_1 & \bar{\phi}_2 & 0 \end{array}\right).
\end{equation}
Unlike the geometries we have used in our examples, the geometry
used by Rovelli does satisfy all the axioms for a real spectral triple.
The eigenvalues and eigenvectors of $D+JDJ^{-1}$ are
\begin{eqnarray}
\lambda = 0 & : &\left(\begin{array}{ccc}
|\phi_2|^2 & -\phi_1 \bar{\phi}_2 & 0 \\
-\bar{\phi}_1 \phi_2 & |\phi_1|^2 & 0 \\
0 & 0 & 0 \end{array}\right), \\
\lambda = \pm\frac{2}{\hbar}\sqrt{|\phi_1|^2+|\phi_2|^2} & : &\left(\begin{array}{ccc}
|\phi_1|^2 & \phi_1 \bar{\phi}_2 & \pm\sqrt{|\phi_1|^2+|\phi_2|^2}\,\phi_1 \\
\bar{\phi}_1 \phi_2 & |\phi_2|^2 & \pm\sqrt{|\phi_1|^2+|\phi_2|^2}\,\phi_2 \\
\pm\sqrt{|\phi_1|^2+|\phi_2|^2}\,\bar{\phi}_1 & \pm\sqrt{|\phi_1|^2+|\phi_2|^2}\,\bar{\phi}_2 & |\phi_1|^2+|\phi_2|^2 \end{array}\right),
\end{eqnarray}
so
\begin{equation}
S = \frac{8\LPlanck^2}{\hbar}\left(|\phi_1|^2+|\phi_2|^2\right).
\end{equation}
This has a U(2) symmetry under the
$\Inn(\algebra) \cong \Ugrp(2)\times\Ugrp(1)$
gauge transformations
\begin{equation}
D \to (uJuJ^{-1})D(uJuJ^{-1})^\dagger
= D+u[D,u^\dagger]+Ju[D,u^\dagger]J^{-1}.
\end{equation}
An overall factor of U(1) acts trivally because it commutes with $D$.

Quantising the action, we get the gauge-fixed partition function
\begin{equation}
Z = \int\limits_0^\infty\!\d\phi\, \phi^3\exp\left(-\frac{8\phi^2}{\MPlanck^2}\right)
= \frac{\MPlanck^4}{128},
\end{equation}
where $\phi := \sqrt{|\phi_1|^2+|\phi_2|^2}$.
From this, we find the Greens functions to be
\begin{equation}
\vev{\phi^n}
= \Gamma\left(\frac{\scriptstyle n+4}{\scriptstyle 2}\right)
\left(\frac{\MPlanck}{\sqrt{8}}\right)^n.
\end{equation}
For the distance used in \cite[eqn.~22]{Rovelli:ncg},
\begin{equation}
d = \frac{\hbar}{\sqrt{|\phi_1|^2+|\phi_2|^2}} = \frac{\MPlanck}{\phi}\LPlanck,
\end{equation}
the expectation values are
\begin{equation}
\vev{d}_N = \frac{\Gamma(N+\frac{\scriptstyle 3}{\scriptstyle 2})}{\Gamma(N+2)} \sqrt{8}\,\LPlanck.
\end{equation}

\section{Spectral Integrals}
\label{sec:spectral}

A proposal for a path integral approach is also put forward in \cite{Rovelli:ncg}.
It suggests that the integration measure should be given by the
eigenvalues of the Dirac operator.
This complements the spectral invariance of the spectral action.
We shall refer to such path integrals as \textit{spectral integrals}.

Spectral integrals differ from our path integrals in the way
they integrate over the space of Dirac operators.
The starting point for both is the space of self-adjoint operators,
which can be partitioned into unitary equivalence classes.
In our approach, we quotient out all those operators that have
a non-zero trace, to leave only traceless self-adjoint operators.
We then remove any degrees of freedom
belonging to the center of the $C^*$-algebra $\algebra$.
This has the effect of reducing the unitary equivalence classes
down to $\Inn(\algebra)$ equivalence classes.
The space we are left with is the space of Dirac operators that we
integrate over.
We use gauge-fixing to perform the integration, so path integrals
separate into a contribution from the gauge orbits and an integral
along a section.

In contrast, spectral integrals just integrate over the orbit space of the
unitary group action on the space of self-adjoint operators.
The orbit space has the operator eigenvalues as cartesian coordinates,
so there is no dependence on $\Inn(\algebra)$.
(To be precise, one should order the eigenvalues,
$\lambda_1<\lambda_2<\ldots<\lambda_n$,
or include a symmetry factor in the integrals.)
This means that different finite geometries with representations of the
same dimension will have the same spectral integrals.

As a case in point, take the two-point space and the matrix geometry $\Matrix{2}{\Cset}$.
Both have two-dimensional representations and so two Dirac operator eigenvalues.
Their spectral integrals will therefore be identical, making it impossible
to distinguish between the two geometries using expectation values alone.
For example, both geometries have the distance v.e.v.
\begin{equation}
\vev{d} = \frac{1}{Z}\int\!\d\lambda_1\,\d\lambda_2\, \frac{\sqrt{2}\,\LPlanck^2}{\sqrt{\lambda_1{}^2+\lambda_2{}^2}} \exp\left(-\frac{\lambda_1{}^2+\lambda_2{}^2}{\LPlanck^2}\right)
= \sqrt{2\pi}\,\LPlanck.
\end{equation}

It should be remembered that spectral integrals are, so far, just an idea,
and we have interpreted it literally.
An obvious refinement that could be made is to impose a
traceless condition on the eigenvalues.
We await to see if there are any further developments.

\section{Riemannian Manifolds}
\label{sec:riemann}

We now move on to outline how our approach might work for
less trivial noncommutative geometries, in particular Riemannian
manifolds.
The Dirac operator for a Riemannian manifold is
\begin{equation}
D := -\im\gamma^a e_a^\mu(x)
\left(\pderiv{}{x^\mu}
+\frac{1}{4}\omega_{bc\mu}(x) \gamma^b\gamma^c\right),
\end{equation}
where $e^a_\mu$ is the vierbein and $\omega^{ab}_\mu$ is the
spin connection.
(Note: $\Tr D=0$, as each term contains an odd number of gamma matrices.)
Computing the spectral action yields the Einstein-Hilbert action
(ignoring higher order terms).
The details of the calculation can be found in \cite{Connes:spectral}.

Usually, the metric, $\guv$, is considered as the dynamical field
and hence gives the measure for path integrals.
In our approach, the vierbein and spin connection would be used
instead, these being the degrees of freedom of the Dirac operator.
This resembles the conventional connection-based way of
quantising Yang-Mills theories.
So, one might hope that this will make things more tractable.

We can go further.
Let us now use a Dirac operator with a self-dual
spin connection $A^{ab}_\mu$.
Since we work in an Euclidean signature, $A^{ab}_\mu$ is real as
$A^{ab}_\mu = \half\epsilon^{ab}{}_{cd}A^{cd}_\mu$
(it is complex in a Lorentzian signature).
Applying this constraint to the spectral action will give the
Einstein-Hilbert action with a self-dual curvature.
This is essentially the Ashtekar formulation of general relativity.

The canonical quantisation, with respect to the spin connection,
proceeds by performing a $3+1$ ADM decomposition.
From this, the conjugate momentum, $\pi_{ab}^\mu$, can be determined.
It is self-dual and related to the vierbein.
Making use of the self-duality, one can define the variables
\begin{equation}
A^i_\mu := A^{0i}_\mu, \quad \pi_i^\mu := \pi_{0i}^\mu,
\end{equation}
where $i=1,2,3$ is a space index.
Their Poisson bracket is
\begin{equation}
\{A^i_\mu(x),\pi_j^\nu(y)\} = \delta_\mu^\nu \delta^i_j \delta^3(x-y).
\end{equation}
This is very much like the Yang-Mills situation,
with $i$ and $j$ as the SO(3) group indices.
There are also constraint equations, the most notorious of which,
is the Hamiltonian constraint.
The quantisation of the constraints is dealt with by using
loop representations \cite{Gambini_Pullin}.
This is the origin of loop quantum gravity.

The path integral quantisation is related to spin foams.
It is possible to write the Einstein-Hilbert action in
the form of a $BF$ theory,
\begin{equation}
S := \int_M\!e_a^\mu e_b^\nu F^{ab}_{\mu\nu} e \,\d^4x
= \int_M\!B_{ab}^{\mu\nu} F^{ab}_{\mu\nu}\sqrt{g}\,\d^4x,
\end{equation}
where $F^{ab}_{\mu\nu}$ is the self-dual curvature,
and $B_{ab}^{\mu\nu} = e_a^\mu e_b^\nu$ is a constraint.
Path integrals over the spin connection and vierbein then resemble
the quantisation of $BF$ theory.
To make the path integrals well-defined, they can be discretised
by triangulating the manifold.
In $BF$ theory, this gives rise to the concept of spin foams
\cite{intro-spinfoam}, the spin network equivalent of Feynman diagrams.

\section{Conclusion}
\label{sec:conclusion}

We have developed a path integral approach to quantise the
spectral action.
It has been successfully applied to the two-point space and the
matrix geometry $\Matrix{2}{\Cset}$.
In both cases, graviton excitations have the effect of shrinking distances.
However, the behaviour of the geometries as they collapse to a point
is quite different.
The two-point space undergoes a topological change,
which is suggestive of the formation of something like a black hole
(an apt term would be ``\textit{black point\/}").
Whereas, the matrix geometry maintains its topology,
but loses its noncommutativity instead.
We expect the shrinking of distances by gravitons to be a
general feature of quantised finite noncommutative geometries.
The introduction of fermions onto the geometries had the effect of
shielding out the gravitational field.
All the graviton states are lowered by an amount equal to
the number of fermion generations.

Comparing our approach with \cite{Rovelli:ncg},
led us to question the validity of their results.
We found that their equations of motion could be expressed in much
simpler terms, which result in a smaller phase space.
This will alter their canonical quantisation.
Despite this, both approaches seem to support the qualitative result
that distances shrink with increasing graviton excitations.

The idea of spectral integrals is very appealing as it is
consistent with the philosophy of spectral invariance.
But, we have concerns over the possible lack of any
topological dependence.
The K-groups should somehow determine the (sub)space of
eigenvalues to integrate over.
On Riemannian manifolds, our path integral approach coincides
with the conventional one, by construction.
It would be interesting to see how spectral integrals differ from this.

The next step, to obtaining a better understanding of
quantised noncommutative geometries, would be to
investigate some more substantial examples than the ones we have
considered here.
For example, the spectral triple associated with the finite part of
the standard model $C^*$-algebra,
i.e.\ $\Cset\oplus\Hset\oplus\Matrix{3}{\Cset}$.
We hope to pursue these ideas in the future.

Noncommutative geometry has introduced a new twist in the
search for a theory of quantum gravity.
The biggest problem we face may not be one of quantisation,
but one of finding the right geometry to quantise.

\section{Acknowledgements}

The author wishes to thank David Fairlie and Bill Oxbury for useful
discussions, and PPARC for a studentship award.

\bibliographystyle{preprint}
\bibliography{ncg,qncg,qgravity}

\end{document}